\documentclass[nosumlimits]{amsart}
\usepackage{amsmath}
\usepackage{amssymb}
\usepackage{layout}

\usepackage{graphicx}

\newtheorem{thm}{Theorem}[section]
\newtheorem{defin}[thm]{Definition}

\newtheorem{cor}[thm]{Corollary}

\newtheorem{pro}[thm]{Proposition}
\numberwithin{equation}{section}

\newenvironment{remark}[1]{\medskip\par\noindent\small\
\begin{center}\textbf{Remark}\end{center}\begin{quote} #1}
{\medskip\par\noindent\end{quote}}

\newcommand{\D}{\mathcal{D}}  

 \newcommand{\into}{\rightarrow}

 \newcommand{\set}[1]{\left\{#1\right\}}
 
 \newcommand{\dotp}[2]{\left<#1,#2\right>}

 \newcommand{\norm}[1]{\left\Vert#1\right\Vert}
 
 \newcommand{\qtext}[1]{\quad\text{#1}\quad}
 \newcommand{\fa}{\qtext{for all}}
 \newcommand{\bb}{\begin{equation*}}
 \newcommand{\ee}{\end{equation*}}
 \newcommand{\bp}{\begin{proof}}
 \newcommand{\ep}{\end{proof}}

\begin{document}

\title[Fields generated by flows of matter]{Relativistic gravity fields\\ and electromagnetic
fields\\ generated\\ by flows of matter}

\author{ Victor M. Bogdan }

\address{Department of Mathematics, McMahon Hall 207, CUA, Washington DC 20064, USA}

\date{7 July 2009}

\email{bogdan@cua.edu}


\begin{abstract}
One of the highlight of this note is that
the author presents the relativistic gravity field
that Einstein was looking for.
The field is a byproduct of the matter in motion.
This field can include both the discrete and
continuous components. In free space the waves produced in this
field propagate with velocity of light.

Another highlight is the proof of amended Feynman's
formulas for electromagnetic potentials. This makes the
formulas mathematically complete and precise.

The main result can be stated as follows.
In a fixed Lorentzian frame given is a trajectory $r_2(t,r_0)$ of
flow of matter.
The parameter $r_0$ changes in a compact set $F$ representing the position of the
matter at some initial time $t_0.$
The flow must satisfy certain conditions of regularity.

  Given any signed measure $q(Q)$ of finite variation defined on Borel subsets of $F,$
representing total charge contained in the set $Q\subset F,$
such a flow determines the scalar $\phi$ and
the vector $A$ potentials for a pair $(E,B)$ of fields
satisfying Maxwell's equations and as a byproduct
the field $\varrho,$ representing density of charges, and the field
$j,$ representing the density of currents.
All these fields are represented in terms of generalized functions.
\end{abstract}


\maketitle

\bigskip
\section{Uniqueness of time delay field for flows of matter}
\bigskip

The purpose of this note is to present formulas
for for gravitational and electromagnetic field in terms of potentials
generated by flows of matter. We want to consider initial distributions of
charges and mass that include both continuous and discrete components.

For the sake of simplicity in notation we
select units so that the speed of light is $c=1,$ and the electrostatic
constant satisfies the condition $4\pi\epsilon_0=1,$ and the universal
constant of gravity $G=1.$ We are working here
in a fixed Lorentzian frame.

This means that if we keep meter as our unit of length the unit of time
is approximately 3.3 nanoseconds and the unit of mass is about $3.871$
metric tons and the unit of charge is $4\pi$ of coulombs.

\bigskip

\begin{defin}[Flow of matter]
Let $F\subset R^3$ be a compact set representing the position of matter
at some initial time $t_0.$

By a {\bf flow of matter} we shall understand a continuous function
$(t,r_0)\mapsto r_2(t,r_0)$ from the
product $R\times F$ into $R^3,$ infinitely differentiable with respect to $t$ and
such that the velocity $v_2(t,r_0)=\dot{r}_2(t,r_0)$ and the acceleration
$a_2(t,r_0)=\dot{v}_2(t,r_0)$ and the derivative $\dot{a}_2(t,r_0)$ and so on,
are continuous on the product  $R\times F.$

Moreover the following two conditions are satisfied:
\begin{itemize}
    \item For every time $t_1$ there is a velocity
$v_1<c=1$ such that
\begin{equation}\label{velocity v 1}
    |v_2(t,r_0)|\le v_1\fa t\le t_1\text{ and } r_0\in F.
\end{equation}

    \item  For every time $t\in R$ the map $P_t$ given by the formula
\begin{equation}\label{P t map}
    P_t(r_0)=r_2(t,r_0)\fa t\in R\text{ and } r_0\in F
\end{equation}
represents a one-to-one mapping of $F$ onto $P_t(F).$
\end{itemize}
\end{defin}

\bigskip

Clearly we have
\begin{equation*}
    P_{t_0}(r_0)=r_2(t_0,r_0)=r_0\fa r_0\in F.
\end{equation*}
The function $t\mapsto r_2(t,r_0)$ will be called a {\bf line of flow} corresponding to
the index $r_0.$

\begin{remark}
The above definition of flow of matter depends on the Lorentzian frame.
To make it frame independent define the set $F$ as a subset of a hyperplane
of codimension 3 consisting of space like points.

Define the lines of
flow as time like trajectories of point mass satisfying certain continuity
conditions and conditions on the initial parts of flow to be able
to find the representation of the flow in any Lorentzian frame as
defined above.

We leave the details to the reader.
\end{remark}
\bigskip

Let $T=T(r_1,t,r_0)$ denote the time delay required to reach
point $r_1\in R^3$ at time $t$ from the line of flow corresponding to
index $r_0.$ Its value must satisfy the Lorentz \cite{lorentz}
time delay equation
\begin{equation}\label{time delay}
    T=|r_1-r_2(t-T,r_0)|.
\end{equation}

\begin{thm}[Time delay is unique and continuous]
For every point $r_1\in R^3,$ and time $t\in R,$ and index $r_0\in F$
there exists one and only one solution $T$ of equation (\ref{time delay}).
Moreover the function $T=T(r_1,t,r_0)$ is continuous on its entire domain
$R^3\times R\times F.$
\end{thm}
For a proof of an analogous theorem see \cite{bogdan61}.
Notice the following relations
\bb
    T=0\Leftrightarrow\{ r_1=r_2(t,r_0)\text{ for some }r_0\in F\}
    \Leftrightarrow r_1\in P_t(F).
\ee
The set $P_t(F)$ represents the position of matter at time $t.$

Now define set
$$
    G=\set{(r_1,t,r_0)\in R^3\times R\times F:\ T(r_1,t,r_0)>0}.
$$
The set $G$ represents the natural domain of the field $u=T^{-1}.$

\bigskip
\section{Fundamental fields corresponding to the flow}
\bigskip

\begin{defin}[Fundamental fields]
\label{fundamental fields}
Introduce the retarded time function
\begin{equation*}
 \tau=\tau(r_1,t,r_0)=t-T(r_1,t,r_0)\fa (r_1,t,r_0)\in R^3\times R\times R^3,
\end{equation*}
retarded velocity
\begin{equation*}
 v=v_2(\tau(r_1,t,r_0),r_0)\fa (r_1,t,r_0)\in R^3\times R\times R^3,
\end{equation*}
 and retarded acceleration
\begin{equation*}
 a=a_2(\tau(r_1,t,r_0),r_0)\fa (r_1,t,r_0)\in R^3\times R\times R^3,
\end{equation*}
and vector field $r_{12}$ by
\begin{equation*}
 r_{12}=r_1-r_2(\tau(r_1,t,r_0),r_0)\fa (r_1,t,r_0)\in R^3\times R\times R^3.
\end{equation*}
Introduce the unit vector field $e,$ and the fields $u$ and $z$
by the formulas
\begin{equation}\label{unit vector field}
    \qtext{and}e=\frac{r_{12}}{T}\qtext{and} u=\frac{1}{T}
    \qtext{and}z=\frac{1}{(1-\langle e,v\rangle)}\qtext{on} G.
\end{equation}
These functions will be called the {\bf fundamental fields} associated
with the flow $r_2(t,r_0),$ where $t\in R$ and $r_0\in F.$
\end{defin}

Notice that by definition of flow of matter
the velocities are smaller in magnitude than
the speed of light $c=1.$ Thus we must have for the dot product
$|\langle e,v\rangle|\le|v|<1.$
So the field $z$ is well defined.

All the above functions consist of compositions of continuous functions,
therefore each of them is continuous on its respective  domain and thus
all of them are continuous on their common domain, the set $G.$

\bigskip

We would like to stress here that the fundamental fields depend on the Lorentzian
frame, in which we consider the flow.
It is important to find expressions involving fundamental fields that
yield fields invariant under Lorentzian transformations.

Lorentz and Einstein \cite{einstein2a}, Part II, section 6, established that
fields satisfying Maxwell equations are invariant under Lorentzian transformations.
\bigskip

Our main goal is to prove that fields constructed for flows
of matter will satisfy Maxwell equations.
We shall do this by showing that these fields
are representable by means of fundamental fields and using the formulas
for partial derivatives of the fundamental fields prove that
such fields generate fields satisfying Maxwell equations.

Introduce operators $D=\frac{\partial}{\partial t}$ and
$D_i=\frac{\partial}{\partial x_{i}}$ for $i=1,2,3$ and
$\nabla=(D_1,D_2,D_3).$

Observe that $\delta_i$ in the following formulas denotes the i-th
unit vector of the standard base in $R^3$ that is
$\delta_1=(1,0,0),$ $\delta_2=(0,1,0),$ $\delta_3=(0,0,1).$

The symbols $e_i,$ $v_i,$ $a_i,$ denote the corresponding component
of the vector fields $e,$ $v,$ $a,$ respectively.

\bigskip

\begin{thm}[Partial derivatives of fundamental fields]
Assume that in some Lorentzian frame
we are given a plasma flow $(t,r_0)\mapsto r_2(t,r_0).$
For partial derivatives with respect to coordinates of the vector $r_1$
we have the following identities on the set $G$
\begin{eqnarray}
\label{DiT}       D_iT&=&ze_i  ,\\
\label{Diu}       D_iu&=&-zu^2e_i  ,\\
\label{Div}          D_iv&=&-e_iza ,\\
\label{Di tau}    D_i\tau&=&-ze_i,\\
\label{Die}         D_ie&=& -uze_ie+u\delta_i+uze_iv \qtext{where} \delta_i
                    =(\delta_{ij}),\\
\label{Diz}          D_i z &=& -z^3e_i \langle e,a \rangle
                -uz^3e_i+ uz^2e_i+uz^2v_i+uz^3e_i \langle v,v \rangle \\
\label{grad T}\nabla T&=&ze  ,\\
\label{grad u}\nabla u&=&-zu^2e ,\\
\label{grad z} \nabla z&=&  -z^3 \langle e,a \rangle e-uz^3e+ uz^2e+uz^2v
                +uz^3 \langle v,v \rangle e.
\end{eqnarray}
and for the partial derivative with respect to time we have
\begin{eqnarray}
\label{DT}    DT&=&1-z,\\
\label{Du}    Du&=&zu^2-u^2,\\
\label{D tau}    D\tau&=&z,\\
\label{Dv}    Dv&=&za  ,\\
\label{De}    De&=&-u e+ u z e-u z v,\\
\label{Dz}    Dz&=&uz-2uz^2 +z^3 \langle e,a \rangle +uz^3-uz^3 \langle v,v \rangle .
\end{eqnarray}
Since the expression on the right side of each formula represents
a continuous function, the fundamental fields are at least of class $C^1$ on the set $G.$
Since the lines of flow $t\mapsto r_2(t,r_0)$ of the matter
are of class $C^\infty,$ we can prove by induction that the fundamental
fields are of class $C^\infty$ on $G.$
\end{thm}
%
\bigskip
The proof of the above theorem  is similar to the proof of analogous theorem in
Bogdan \cite{bogdan65}.

\section{Integration with respect to a signed measure}

Let $V$ be a prering of subsets of $F$
consisting of sets of the form $Q\cap B$ where $Q$ is compact and $B$
is open. See Bogdanowicz \cite[page 498]{bogdan10}.

Assume that the set functions $q^+(A)$ and $q^-(A)$ represent, respectively,
the total positive and total negative charge  contained in the body covered by
the set $A\in V.$
We shall assume that these functions are countably
additive.

\begin{remark}
A heuristic argument relying on assumption
that charge of an electron is indivisible can be presented as follows:
Take a decomposition of a set $A\in V$
into a countable union of disjoint sets
\bb
    A=A_1\cup A_2\cup\ldots A_n\cup\ldots.
\ee
Since every charge comes in the form of finite number of indivisible unit charges,
that are all equal to the charge of a single electron,
only a finite number of the sets may contain
a charge. Thus starting from a sufficiently large index $n_0$ all sets $A_n$
will have charge zero. Thus
\bb
    q^+(A)=\sum_{n\le n_0}q^+(A_n)+\sum_{n> n_0}0=\sum_{n=1}^\infty q^+(A_n).
\ee
Similarly we can get countable additivity of $q^-.$
\end{remark}

Put
$q(A)=q^+(A)+q^-(A)$ and $\eta(A)=q^+(A)-q^-(A).$
The value $q(A)$ represents the total charge in the body covered by the
set $A$ and $\eta(A)$
represents a non-negative countably additive set function on $V$
such that $|q(A)|\le \eta(A).$

Such a function satisfies the requirements of a volume function
as defined in \cite[page 492]{bogdan10}. Observe that the
function $q$ belongs to the space $M,$ defined on page 492, and its norm $\norm{q}\le 1.$
Therefore we can use the trilinear integral $\int u(f,dq)$ developed there. In our case
for the bilinear operator $u(y,r)=ry=yr$ defined for $y\in Y$ and $r\in R,$
where $Y$ stands for either the vector space $R^3$ or the space of $R$ of reals.

Thus we can use the theory
developed in the papers Bogdanowicz \cite{bogdan10} and \cite{bogdan14}. Both
papers are available on the web.

The classical theory of measure based on sigma rings of sets, and the Lebesgue theory
of integration and theory of Bochner integral follow from
these two papers, making all the classical tools of measure and integration
available if needed in applications.

\bigskip
Concerning notation: We are using the symbol $\int u( f,dq)$ to denote the integral
over the entire space $F$ of integration. When it is desirable to indicate the variable
of integration we shall write $\int u(f(r_0),q(dr_0)).$

If we have a set $A\subset F$
and a function $f:F\mapsto Y$ such that the product $\chi_Af$ yields an
$\eta$-summable function,
where $\chi_A$ denotes the characteristic function of the set $A,$
then we shall say that the function $f$ is summable on the set $A$ and by its
integral over the set we shall understand the following
\begin{equation*}
    \int_A u(f,dq)=\int u(\chi_Af,dq).
\end{equation*}
Since $\chi_Ff=f$ for all functions defined on $F,$
the two notions for the set $F$ coincide,  that is
\begin{equation*}
    \int u(f,dq)=\int_F u(f,dq).
\end{equation*}

In the case when the bilinear is of the form $u(r,\lambda)=r\lambda=\lambda r,$
where $r$ is a vector and $\lambda$ is a scalar, we shall
write the integral with respect to $u$ just as $\int f\,dq.$



\section{Basic notions from the theory of generalized vector fields}
\bigskip


It will be convenient here to use the theory of generalized functions, called
also distributions,
originally introduced heuristically by Dirac in his works
on Quantum Mechanics and put on precise mathematical footing by
L. Schwartz \cite{schwartz} and Gelfand
and Shilov \cite{gelfand1}.

Assume that $G$ represents an open set in $R^4.$
Let $\D_k$ denote the space $C^\infty_0(G,R^k)$
of infinitely differentiable functions having compact supports contained in $G$ and
values in $R^k.$

\begin{defin}[Sequential topology on $\D_k$]
On the space $\D_k$ we introduce a {\bf sequential topology.} We shall say
that a sequence
of functions $g_n\in \D_k$ converges to a function $g\in \D_k$ in
the sequential topology if it
converges uniformly together with all its partial derivatives $D^\alpha g$,
on every compact subset $K$ of $G,$ to the function $g,$ that is for every $\alpha $
the following sequence converges uniformly on $ K $
$$
    D^\alpha g_n(x)\into D^\alpha g(x)
$$
where $\alpha=(k_1,\cdots,k_4)$ denotes the multi-index of a partial derivative
$$D^\alpha=D_1^{k_1}\cdots D_4^{k_4}$$
with
$k_j=0,1,2,\ldots$ and $D_j=\frac{\partial}{\partial x_j}$ where $j=1,2,3,4.$
\end{defin}
\bigskip

\begin{defin}[Generalized vector valued functions]
\label{Generalized vector valued functions}
Let $\D'_k$ denote the space of all linear continuous real functionals on $\D_k.$
Convergence in $\D'_k$ will be understood as pointwise convergence. The space
$\D'_k$ will be called the space of {\bf generalized vector valued functions} or
the space of vector distributions.
\end{defin}

We shall use the following equivalent notation for such a functional
\begin{equation}\label{integral formula for distribution}
    f(g)=\dotp{f}{g}=\int_G f(x)\cdot g(x)\,dx\fa g\in \D_k.
\end{equation}
In the above  $y\cdot x$
denotes the dot product, also called the scalar product, of two vectors $y,x\in R^k.$

Any continuous function $f$ on the set $G$ generates by means of the above integral
formula a vector distribution.

Notice that
the space $\D'_k$ is linearly and topologically isomorphic with the Cartesian
product of $k$ copies of the space $\D'_1$
$$
    \D'_1\times\cdots\times\D'_1=(\D'_1)^k.
$$

Indeed, if for every $m=1,\dots,k$ we define maps $P_m:R\into R^k$ by the condition
\begin{equation*}
    P_m(t)=x\Leftrightarrow \{x_m=t\text{ and }x_j=0\text{ if }j\not=m\},
\end{equation*}
where $x=(x_1,\ldots,x_k)\in R^k,$
then the functionals
\begin{equation*}
    f_m(g)=\int_G f(x)\cdot P_m(g(x))\,dx\fa g\in \D_1
\end{equation*}
are well defined and represent elements of the space $\D'_1.$
Thus the element
\begin{equation*}
 (f_1,\cdots,f_k)\in (\D'_1)^k.
\end{equation*}

Conversely define maps $P'_m:R^k\into R$ for $m=1,\ldots,k$ by the formula
\begin{equation*}
    P'_m(x)=x_m\fa x\in R^k,\ m=1,\ldots,k.
\end{equation*}
Then the  formula
\begin{equation*}
    f(g)=\sum_m\int_G f_m P'_m(g(x))\,dx\fa g\in \D_k
\end{equation*}
yields a linear continuous functional on the space $\D_k.$
It is easy to verify that the transformation $Q:\D'_k\into (\D'_1)^k$defined by
\begin{equation*}
    f\mapsto (f_1,\ldots,f_k)
\end{equation*}
is indeed a linear and topological isomorphism of the two spaces.
\bigskip

Now notice the following fact.

\begin{pro}[Imbedding of continuous functions into $\D'_k$ is one-to-one]
Given two functions $f_1$ and $f_2.$ Assume that they are continuous on
the open set $G$ with exception perhaps of points lying on an admissible trajectory
of a point mass.
If they generate the same
vector distribution, that is
\begin{equation*}
    \int_G f_1(x)g(x)\,dx=\int_G f_2(x)g(x)\,dx\fa g\in \D_k,
\end{equation*}
then they coincide
\begin{equation*}
    f_1(x)=f_2(x)
\end{equation*}
at every point $x\in G$ with exception perhaps of the points on the trajectory.
\end{pro}
\bigskip

\bp
Indeed, from linearity of the map $f\mapsto \dotp{f}{g}$ and the previous isomorphism
it is sufficient to prove that for any real function $f$ continuous
at every point except perhaps points lying on the admissible trajectory
such that
\begin{equation*}
    \int_G f(x)g(x)\,dx=0\fa g\in \D_1
\end{equation*}
follows that $f=0$ on $G$ with the exception of points on the trajectory.
\bigskip

Assuming that this is not true
then at some point $x_0\in G$ we have $f(x_0)\neq 0.$ We may assume without loss of
generality that $2\delta=f(x_0)>0$ otherwise we would consider the function $-f.$
From continuity of $f$ follows that there is a rectangular
neighborhood $V\subset G$ of $x_0$ such that
\begin{equation*}
    f(x)\ge \delta \fa x\in V.
\end{equation*}

There exists a nonnegative function $g$ of class $C^{\infty}$ with
support in the set $V$ with integral $\int_G g(x)\,dx=1.$ Thus
for such a function we would get
\begin{equation*}
    \int_G f(x)g(x)\,dx=\int_V f(x)g(x)\,dx\ge \int_V \delta g(x)\,dx=\delta>0.
\end{equation*}
A contradiction. So all we have to do is to show that there exists
a function $g$ having the above properties. To this end
consider a nonnegative function $g_0(t)$ defined by the formula
\begin{equation*}
    g_0(t)=\alpha e^{-1/(1-t^2)}\text{ if }|t|<1;\qtext{and}g_0(t)=0\text{ if }|t|\ge 1,
\end{equation*}
where the constant is selected so that $\int_{-1}^1g_0(t)\,dt=1.$
It follows from the above formula that the function $g_0$ is
infinitely differentiable at every point except perhaps at $t=+1$ or $t=-1.$
One can prove that at these points the one-sided derivatives exist and
they are equal. Thus the function $g_0$ is of class $C^\infty.$

Now consider the function
\begin{equation*}
    g_1(x)=g_0(x_1)g_0(x_3)g_0(x_2)g_0(x_4)\fa x=(x_1,x_2,x_3,x_4)\in R^4
\end{equation*}
with the integral, over the cube in $R^4$ representing its support, equal to 1.
For the fixed $x_0\in G$ and sufficiently large $n$ we see that the function
given by the formula
\begin{equation*}
    g_n(x)=n^4g(n(x-x_0))\fa x\in R^4
\end{equation*}
will satisfy our requirements.
\ep

Any linear continuous operator $H :\D_k\into\D_k$ generates a linear continuous
dual operator $H ':\D'_k\into\D'_k$ by the formula
\begin{equation*}
    \dotp{H 'f}{g}=\dotp{f}{H \, g}\fa g\in \D_k.
\end{equation*}

The dual operator corresponding to scalar multiplication $g\mapsto \lambda\, g$
is scalar multiplication  $f\mapsto \lambda\, f$ as follows from
the above definition.

\begin{defin}[Generalized differential operator]
\label{generalized partial derivative}
By a {\bf generalized partial derivative} $D_i=\frac{\partial}{\partial x_i}$
acting onto the m-th component of $f\in \D'_k$
\begin{equation*}
    D_{i,m}(f_1,\ldots, f_k)=(f_1,\ldots,D_i f_m,\ldots,f_k)
\end{equation*}
we shall understand the dual operator to the operator acting onto
the m-th component of $g\in\D_k$ by the formula
$$
    (-1)D_{i,m}(g_1,\ldots, g_k)=(g_1,\ldots,(-1)D_i g_m,\ldots,g_k).
$$
\end{defin}
\bigskip

In the case when the set $G$ represents a Cartesian product of open
bounded intervals and the vector function $f$ is continuous together with
$D_if_m,$
the above formula can be easily verified through iterated integral and
integration by parts.
In this case the ordinary partial derivative
will produce the derivative in the sense of distributions.
So it is natural to extend this property to vector distributions.
\bigskip

\begin{defin}[Weak and strong partial derivatives]
\label{def-weak and strong partial derivatives}
Now every distribution and in particular every continuous function is
differentiable in the sense of distributions. If it happens that the
distributional partial derivative is representable by means of a continuous
function, such a function is called a {\bf weak derivative.}
If a vector function
has a continuous partial derivative such a derivative is called
a {\bf strong derivative.}
\end{defin}
\bigskip

It is not obvious that weak and strong derivatives coincide in
the general case of an arbitrary open set $G$ in $R^4.$
This calls for the following theorem.
\bigskip

\begin{thm}[Weak and strong partial derivatives coincide]
\label{weak and strong derivatives coincide}
Assume that $f=(f_1,\ldots,f_k)$ is a vector-valued function on
an open set $G\subset R^4$ and that its $m$-th component
has a continuous partial derivative $D_if_m.$

Then this derivative coincides with the derivative in the sense of distribution,
that is the weak derivative coincides with the strong one.
\end{thm}
\bigskip

\bp
To prove this fact for general open sets in $R^4$
notice that for every point $x\in G$ there exists a neighborhood $V(x)\subset G$ in the form
of the Cartesian product of open intervals.
Restricting our test functions
to functions with support in $V(x)$ will yield that on such a domain
the weak and strong partial derivatives coincide.

Since every open set in
$R^4$ is a union of a countable number of compact sets, from the collection
$V(x)\,(x\in G)$ one can extract a locally finite countable cover of the set $G.$

Now using partition of unity theorem,
(for reference concerning this theorem
see for instance Gelfand and Shilov \cite{gelfand1},
vol. 1, Appendix to Chapter 1, Section 2,)
we can prove this theorem for any open set $G\subset R^4.$
\ep
\bigskip

\section{Wave with gauge equations imply Maxwell equations
for generalized vector fields}
\bigskip

\begin{thm}[Wave and  gauge imply Maxwell equations]
\label{wave-->Maxwell}
Let on the set $R^4$
be given two generalized scalar fields $\phi$ and
$S$ and two generalized vector fields $A$ and $J.$

If these fields satisfy the following wave equations
\begin{equation*}
    \nabla^2\phi-\frac{\partial^2}{\partial t^2}\phi=-S,
    \quad\nabla^2A-\frac{\partial^2}{\partial t^2}A=-J,
\end{equation*}
with Lorentz gauge formula
\begin{equation*}
    \quad \nabla\cdot A+\frac{\partial}{\partial t}\phi=0
\end{equation*}
then the generalized fields $E$ and $B$ defined by the formulas
\begin{equation*}
    E=-\nabla \phi - \frac{\partial}{\partial t} A\qtext{and}B
    =\nabla\times A
\end{equation*}
will satisfy the following Maxwell equations
\begin{equation}\label{Maxwell equations for potentials}
\begin{split}
    &(a)\quad\nabla\cdot E=S,\ \quad(b)\quad\nabla\times E
    =- \frac{\partial}{\partial t} B,\ \quad\\
    &(c)\quad
    \nabla\cdot B=0,\ \quad(d)\quad\nabla\times B
    =\frac{\partial}{\partial t}E+J,
\end{split}
\end{equation}
and the equation of continuity
\begin{equation*}
     DS+\nabla\cdot J=0.
\end{equation*}
\end{thm}
\bigskip


\begin{thm}[Continuity on parameter $r_0$]
\label{wave equations}
For any  flow of matter $r_2(t,r_0)$ the functions
\begin{equation*}
 r_0\mapsto uz\qtext{and}r_0\mapsto uzv
\end{equation*}
are continuous from the set $F$ into the space $\D'_1$ and $\D'_3,$
respectively.
\end{thm}

\bigskip


\begin{thm}[Commutativity of differential and integral operators]
\label{Commutativity of differential and integral operators}
Assume that $r_0\mapsto h$ is a continuous function from
the set $F$ into the space $\D'_k$ of generalized functions on the
open set $G.$

Then the generalized function
\bb
    H=\int_F h(r_0)\,q(dr_0)
\ee
is well defined and we have the following formulas
\begin{equation*}
     D\int_F h\,dq=\int_F Dh\,dq\qtext{and}D_i\int_F h\,dq=\int_F D_i h\,dq.
\end{equation*}
\end{thm}
\bigskip

\begin{defin}[Scalar and vector potentials]
For any flow of matter and any countably additive measure $q(Q),$
where $q(Q)$ represents the charge contained in the space covered by the set $Q\subset F,$
define the {\bf scalar potential} $\phi,$ and
the {\bf vector potential} $A,$
by the formulas
\begin{equation}
    \begin{split}
        \phi(r_1,t)&=\int_F [(uz)(r_1,t,r_0)]\,q(dr_0),\\
        A(r_1,t)&=\int_F [(uzv)(r_1,t,r_0)]\,q(dr_0),\\
    \end{split}
\end{equation}
The functions under the integral are treated as generalized functions
of variable $(r_1,t).$
\end{defin}
\bigskip

\begin{thm}[Potentials are well defined]
For any flow of matter and any measure $q(Q)$ over $F$
the scalar and vector potentials are well defined
as generalized functions of $(r_1,t)\in R^4.$
\end{thm}
\bigskip

Introduce the D'Alembertian operator by the formula
\begin{equation*}
 \Box^2=\nabla^2-D^2.
\end{equation*}

\begin{thm}[Potentials provide solution to Maxwell's equations]
For any flow of matter and any measure $q(Q)$ of finite variation over $F$
define generalized fields $S$ and $J$ by the formula
\begin{equation}
S=-\Box^2 \phi\qtext{and}J=-\Box^2 A
\end{equation}

Then the fields defined by
$E=-\nabla\phi-\frac{\partial}{\partial t}A$ and $B=\nabla\times A$
will satisfy the Maxwell equations
\begin{equation}
    \nabla\cdot E=S,\ \quad\nabla\times E
    =-\frac{\partial}{\partial t}B,\ \quad
    \nabla\cdot B=0,\ \quad\nabla\times B
    =\frac{\partial}{\partial t}E+J
\end{equation}
and can be represented by means of the integral formulas
\begin{equation}\label{feynman's formula}
    E=\int_F \left(u^2e+u^{-1}\frac{\partial}{\partial t}(u^2e)
    +\frac{\partial^2}{\partial t^2}e\right)\,dq,
\end{equation}
\begin{equation*}
     B=\int_F e\times \left(u^2e+u^{-1}\frac{\partial}{\partial t}(u^2e)
    +\frac{\partial^2}{\partial t^2}e\right)\,dq,
\end{equation*}

Moreover the field $S$ represents the generalized density of charges and
the field $J$ represents the generalized density of currents.
They satisfy the equation of continuity
\begin{equation*}
 \nabla\cdot J+\frac{\partial}{\partial t}S=0
\end{equation*}
of flow of charge.
Here $F$ represents the initial position of the plasma in $R^3$,
and the scalar field $u$ and
the vector field $e$ are defined in formula (\ref{unit vector field}).
\end{thm}

\section{Einstein's illusive gravity field}
\bigskip

Einstein using general theory of relativity proved that waves in the
gravity field have to propagate with velocity of light. For reference
see Einstein and Rosen \cite{einstein4}. Since in any field satisfying
the homogenous wave equation, waves propagate with velocity of light,
the fields $E$ and $B$ in free space have this property.
The field should represent the dynamics of matter and should have
both continuous and discrete parts. Clearly the fields we just
investigated satisfy these conditions.

For fields which represent purely discrete fields we have not
only their form but also the dynamics of their evolution in the
form of n-body problems. For references in this regard see Bogdan \cite{bogdan61},
\cite{bogdan64}, and \cite{bogdan65}.

Now we shall consider the gravity field. Assume that at the time $t=t_0,$
we know the measure $m_0(Q)$ representing the distribution of the rest mass
of the system.

Since from the previous considerations follows that the Lorentzian
frame with the formula of the flow of mass determine the potentials
$\phi$ and $A$ and the fields $E$ and $B,$ the physical nature of
the fields is of no importance. Properties of these fields we derived just from
the geometrical part of the nature of the special theory of relativity.

So the gravity field also should be representable by means
of these potentials.
Again, we remind the reader that
we are working in a Lorentzian frame with units
selected so that the speed of light $c=1$ the electrostatic constant
in free space satisfies the condition $4\pi\epsilon_0=1$ and the unit
of mass is selected so that the gravitational constant $G=1.$

This means that if we keep meter as our unit of length the unit of time
is approximately 3.3 nanoseconds and the unit of mass is about $3.871$
metric tons and the unit of charge is $4\pi$ of coulombs.

\begin{defin}[Generalized gravity field]
Let $(t,r_0)\mapsto r_2(t,r_0)$ represent a flow of matter and
$m_0(Q)$ represent the rest mass of the part of space covered
by the set $Q\subset F$ at time $t=t_0.$

Then the intensity $E$ of the gravity field is given
by the formula
\begin{equation*}
 E(r_1,t)=\int_F [\nabla \phi+\frac{\partial}{\partial t}\phi](r_1,t,r_0)\,m_0(dr_0)
\end{equation*}
where all the the fields represent generalized fields.
The associated field $B$ is given by the formula
\begin{equation*}
 B(r_1,t)=-\int_F [e\times(\nabla\times A)](r_1,t,r_0)\,m_0(dr_0)
\end{equation*}
\end{defin}
\bigskip

The pair of fields $(E,B)$ should be treated as one relativistic entity since
Lorentz \cite{lorentz}
and, independently, Einstein \cite{einstein2a}, see Part 2, section 6, have established that
Maxwell equations are invariant under Lorentzian transformations.
To be more precise
a pair of fields $$E=(E_1,E_2,E_3)\qtext{and}B=(B_1,B_2,B_3)$$
that satisfies Maxwell equations transforms
as a part of an antisymmetric
tensor of second rank. The matrix
of this tensor looks as follows
\begin{equation*}
    \begin{bmatrix}
      0 & +E_1 & +E_2 & +E_3 \\
      -E_1 & 0 & +B_3 & -B_2 \\
      -E_2 & -B_3 & 0 & +B_1 \\
      -E_3 & +B_2 & -B_1 & 0 \\
    \end{bmatrix}
\end{equation*}
\bigskip

\section{Completing Feynman formulas for potentials of the electromagnetic field}
\bigskip

Again we are working in a Lorentzian frame in which we have given
a flow $(t,r_0)\mapsto r_2(t,r_0)$ of matter. The fields
$T,$ $r_{12},$ $e,$ $v$ are the fundamental fields
(\ref{fundamental fields}) associated
with the flow.

We consider the case when the measure $q(Q)$  has Lebesgue summable
density $\varrho(r_0).$ In this case
since the field $uz$ is
\begin{equation*}
 uz=\frac{1}{T}\frac{1}{(1-\dotp{e}{v})}=\frac{1}{(|r_{12}|-\dotp{r_{12}}{v})}
\end{equation*}
the potential fields $\phi$ and $A$
have the following form
\begin{equation}
    \begin{split}
        \phi(r_1,t)&=\int_F
        [\frac{1}{(|r_{12}|-\dotp{r_{12}}{v})}(r_1,t,r_0)]\,\varrho(r_0)\,dr_0,\\
        A(r_1,t)&=\int_F
        [\frac{v}{(|r_{12}|-\dotp{r_{12}}{v})}(r_1,t,r_0)]\,\varrho(r_0)\,dr_0.\\
    \end{split}
\end{equation}
The above formulas complete the formulas for potentials
of the electromagnetic field obtained by Feynman by a heuristic argument.
For details see Feynman-Leighton-Sands \cite{feyn}, vol. 2, chapter 15, page 15.15.

Similar representation is valid for the fields corresponding to gravity.

\bigskip


\section{The independence of the fields from initial measure}
\bigskip
Assume that $q(Q)$ represents as before the total charge contained in the body covered by a
set $Q\subset F$ at time $t_0.$ Assume that at some later time $\tilde{t}_0$
the position of the matter is in the set $\tilde{F}$ and
\begin{equation*}
 P(r)=r_2(\tilde{t}_0,r)\fa r\in F
\end{equation*}
represents transformation of points in $F$ at time $t_0$ to points in $\tilde{F}$
at time $\tilde{t}_0.$ Since by definition of a flow of matter the
transformation $P$ is homeomorphism the set $\tilde{F}$ is compact since $F$ is such.

Let $\tilde{V}$ be the prering consisting of intersections of compact
sets with open sets of the space $\tilde{F}.$ Sets of this prering
can be represented as set differences of two compact sets.
Define set function
\begin{equation*}
 \tilde{q}(\tilde{Q})=q(P^{-1}(\tilde{Q}))\fa \tilde{Q}\in \tilde{F}.
\end{equation*}
Since the transformation $P^{-1}$ preserves compact sets and set differences,
the set function $\tilde{q}$ is well defined. We shall prove that
it represents distribution of charges at time $\tilde{t}_0.$

The following theorem shows that the formulas for potentials in
integral form do not depend on transition from one initial time $t_0$
to a later time $\tilde{t}_0.$

\bigskip

\begin{thm}
Let $h(r)$ be a continuous function on the set $F$ with values the
space  $\D'_k$ generalized vector fields.
Let
\begin{equation*}
    \tilde{h}(r)=h(P^{-1}(r))\fa r\in \tilde{F}
\end{equation*}
Then we have the equality
\begin{equation}\label{initial time transition}
 \int_{\tilde{F}} \tilde{h} \,d\tilde{q}=\int_F h \,dq.
\end{equation}
\end{thm}

\bigskip

\begin{cor}
The scalar potential $\phi$ and the vector potential $A$ are independent
of the initial time $t_0$ when the distribution $q$ of charges was observed,
and as a consequence the field $E$ and the  field $B$ also
do not depend on the initial time when the distribution was observed.
\end{cor}

\bigskip


\end{document}